\documentclass[letter,oldversion]{aa}
\usepackage{graphicx}
\usepackage{natbib}
\usepackage{txfonts}

\begin{document}
\title{Accretion rate and the occurrence of multi-peaked X-ray bursts}
\titlerunning{Accretion rate and multi-peaked bursts}
\author{Anna L. Watts \and Immanuel Maurer}
\institute{Max Planck Institut f\"ur Astrophysik,
  Karl-Schwarzschild-Str. 1, 85741 Garching, Germany}
\offprints{A.L.Watts, \email{anna@mpa-garching.mpg.de}}

\date{}
\abstract{Most Type I X-ray bursts from accreting neutron stars have a
  lightcurve with a single peak, but there is a rare population of
faint bursts that are double or even triple peaked. Suggested mechanisms include
polar ignition with equatorial stalling, or multi-step energy release;
the latter being caused by
hydrodynamic instabilities or waiting points in the nuclear reaction
sequence.  We present an analysis of the accretion rate dependence of
the multi-peak bursts, and discuss the consequences for the various
models.  The observations pose particular challenges for the polar
ignition mechanism given current models of ignition latitude
dependence.} 
 
\maketitle

\section{Introduction}

Neutron stars in Low Mass X-ray binaries (LMXBs) accrete hydrogen (H)
or helium (He)
rich material from their
companion stars via Roche lobe overflow.  This accumulating fuel
burns and under certain circumstances
the burning process becomes unstable, resulting in a thermonuclear
flash known as a Type I X-ray burst (for recent reviews see \citet{str06, gal07}).  The nature of the burning depends primarily on accretion
rate \citep{fuj81, fus87, bil98, nar03, coo06}.  At the lowest
accretion rates, bursts are triggered by unstable H burning. Above
$\sim 1$\% of the Eddington 
accretion rate ($\dot{M}_\mathrm{Edd} \sim 1\times 10^{-8}
M_\odot\mathrm{yr}^{-1}$), H burning stabilises.  A shell of He builds
up and this triggers pure He bursts.  Above $\sim$
10\% $\dot{M}_\mathrm{Edd}$, not all of the H can burn before the He
triggers, leading to various classes of mixed H/He bursts. Eventually, above $\sim 30$\%
$\dot{M}_\mathrm{Edd}$, He burning also stabilises and no more bursts are
seen.  Observations support the existence of this upper limit
\citep{cor03, gal07}.    

The vast majority of X-ray bursts have a single peak in the
lightcurve. However, there are some with a multi-peaked structure.
These fall into two classes.  The first are
Photospheric Radius Expansion (PRE) bursts, where the flux reaches the
Eddington luminosity. As the envelope is lifted from the star the
temperature drops and the energy of the emitted photons falls below
the X-ray band, leading to an apparent drop in countrate.  As the
envelope contracts again the temperature rises, and there is a second
peak in X-ray emission \citep{pac83}.  If one looks at the bolometric
rather than 
the X-ray luminosity, the double-peak structure vanishes.  

\begin{figure}
  \resizebox{\hsize}{!}{\includegraphics{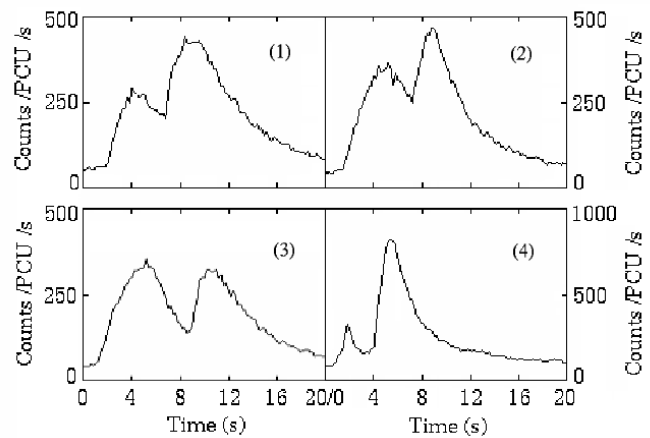}}
  \caption{2-25 keV X-ray lightcurves for the four non-PRE multi-peaked bursts
  in the RXTE archive for 4U 1636-536. Burst numbers (1)-(4) are burst
  numbers 40, 48, 56 and 111 respectively in the burst catalogue of
  \citet{gal07}. PCU refers to the RXTE Proportional Counter Units.   The burst modelled in
  \citet{bha06a} is burst (3); that in \citet{bha06b} is burst (4).}
  \label{f1} 
\end{figure}

In this Letter we focus on the second class of multi-peaked
bursts.  These bursts have lower peak countrates and do not involve
PRE, so that the multi-peak structure is also visible in the bolometric
luminosity. Double-peaked bursts have been reported for 4U 1608-52
\citep{pen89}, GX 17+2 \citep{kuu02} and 4U 1709-267 \citep{jon04}.
The largest sample, however, comes from 4U 1636-536.  Observations
with EXOSAT revealed 5 
double-peaked bursts and one triple-peaked burst
\citep{szt85, van86, lew87}.  Subsequent observations with the {\it Rossi X-ray
  Timing Explorer} (RXTE) have added four more double-peaked
bursts (Figure \ref{f1}, see also \citet{gal07}).  

Since the discovery of these bursts, a number of causes have been
suggested.  \citet{fis04} give an excellent
review of the suggested processes and the observational results that
have led to some being ruled out.  The most likely mechanisms that remain viable
involve a stepped 
release of thermonuclear energy.  This could be caused by
hydrodynamical instabilities leading to mixing \citep{fuj88} or, as
suggested by \citet{fis04}, a waiting point that impedes the nuclear
reaction sequence.  

More recent work has suggested that the critical factor may be the
latitude at which the 
burst ignites.  As will be discussed in more detail below, burst
ignition is most likely to take place on the stellar equator.
\citet{bha06a, bha06b} conjectured that rare 
high latitude ignition could lead to a double-peaked burst if the burning
front slows and partially stalls as it approaches the equator.  This
leads to a dip in the lightcurve:  the second peak forms as
the burning front recovers and accelerates again.  The
authors illustrated the concept by modelling two of the
double-peaked bursts of 4U 1636-536. A particularly attractive feature
of this model is that it attempts to explain simultaneously both the
spectral evolution and 
the detectability 
of millisecond oscillations during the burst rise.  

So what determines ignition latitude?  The general assumption is that accreting material
 comes into corotation and 
hydrostatic equilibrium on a timescale short compared to the
recurrence time between bursts.  The material must therefore spread to
form an equipotential surface that on a rapidly rotating neutron star is
oblate, since the rotation acts to reduce the effective gravity
at the equator.  This means that the {\it local} accretion rate varies with
latitude, being higher at the equator.  \citet{spi02} argued that this
favoured equatorial ignition since it is here that the column depth
necessary to 
trigger a burst is reached most quickly. 

The issue of ignition latitude has now been revisited in more detail
by \citet{coo07} (hereafter CN07). These authors start by assessing the
implications of a local accretion rate above which burning stabilises
\citep{bil98}. As global accretion rate rises, there must come a point
where the equator
has stabilised but higher latitudes have not. In this case,
ignition will occur off-equator.  The situation is complicated by the
existence of different regimes of unstable burning, but detailed
modelling by CN07 confirms the general principle.  Ignition occurs at the equator until a 
certain critical accretion rate (the precise value of which depends on
the stellar  
parameters and the accreted composition).  Above this rate, however,
the ignition point moves rapidly towards the pole and remains there
until the global accretion rate is so high that all bursting activity is
suppressed.    

The accretion rate dependence of the multi-peak bursts may therefore
enable us to distinguish between the various mechanisms.  If the
multi-peaked bursts are indeed caused by polar ignition, for 
example, the results of CN07 indicate that we should expect to
find them at higher global accretion rates than the single-peaked
bursts.  In Section \ref{data} we present an analysis of the accretion rate
dependence of the multi-peak bursts. The implications for the various
models are discussed in Section \ref{models}.  
  
\section{Observations and Data Analysis}
\label{data}

To test the accretion rate dependence of the multi-peak bursts seen
with RXTE we can make use of the RXTE burst catalogue \citep{gal07},
hereafter G07.  This
catalogue records, for over 1000 bursts, a complete set of burst
parameters including two that can be used to assess accretion
rate. The first is position in the colour-colour diagram.  Sources
like 4U 1636-536, known as atoll sources, trace out a path in the 
colour-colour diagram \citep{has89}.  One can assign a colour-colour coordinate $S_z$ that
traces progress though the diagram and is thought to be
proportional to global accretion rate (Figure 4 of G07)\footnote{For
  atoll sources the colour-colour coordinate is usually called $S_a$
  rather than $S_z$ (which is used for the higher accretion rate Z sources). In this paper, however, we use $S_z$ for consistency with G07.}.  A precise calibration of this
relationship, however, is difficult.  The second measure that is used
to estimate accretion rate is the persistent
luminosity.  This is considered to be less reliable than
$S_z$, but its advantage is that it can be used to calibrate the
accretion rate as a fraction of $\dot{M}_\mathrm{Edd}$ provided that the
distance is known (from, for example, a PRE burst). For a full
discussion of these measures we refer the reader
to Section 2.1 of G07.  

\begin{figure}
  \resizebox{\hsize}{!}{\includegraphics{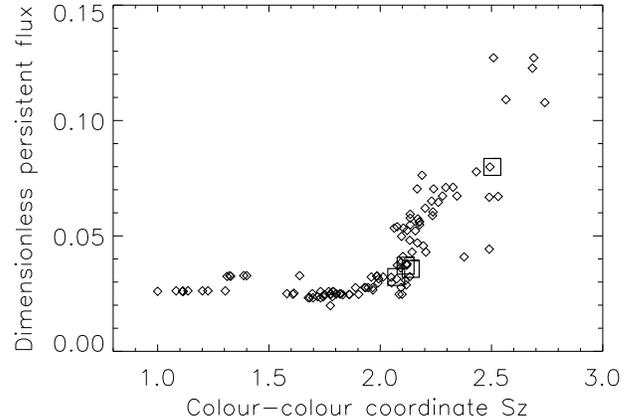}}
  \caption{Dimensionless persistent flux (as a fraction of the
    Eddington rate) against colour-colour diagram coordinate $S_z$ for
    X-ray bursts from 4U 1636-536, data from the RXTE burst catalogue
    (G07).  $S_z$ 
    is assumed to be proportional to accretion rate.   Single peaked
    bursts are shown as diamonds, 
  and the four multi-peaked non-PRE bursts by squares.  Bursts (1)-(3) from Figure \ref{f1} cluster
  together; Burst (4) sits alone.}
  \label{f2}
\end{figure}

We start by considering 4U 1636-536, the source with the largest
sample of multi-peak bursts.  The RXTE catalogue contains 123 bursts
from this source.  Using PRE
bursts as standard candles, G07 estimate that bursts are seen for accretion rates in the range 2-13\%
$\dot{M}_\mathrm{Edd}$.   Figure \ref{f2} shows the distribution of
single-peaked and multi-peaked bursts in a plot of
persistent flux (normalised by the Eddington flux) against
$S_z$.  

Figure \ref{f2} shows that the multi-peaked bursts do not
appear at the highest inferred accretion rates.  There are, even for
the most extreme burst (burst number (4) in Figure \ref{f1})
single-peaked bursts at higher accretion rates.  The EXOSAT data for
4U 1636-536 support these findings.  In May 1984 
\citet{szt85} made a day-long observation during which the persistent
flux declined by $\approx 30$ \%.  Eight bursts were detected, four of
which were 
double-peak non PRE bursts.  Although the bursts
were seen when the flux was relatively high, two single-peak bursts
were also detected at this time.  Later analysis by \citet{pri97}
showed that at this time the source was in the middle banana state
($2.0< S_z< 2.5$), not the higher accretion rate upper banana state.
The later double and triple-peaked bursts reported by \citet{van86}
and \citet{lew87} were also seen in a sequence of over 20 other bursts, at a time
when the flux was rising.  Although both were seen towards the higher
end of the observed flux, there are other bursts in the sequence at higher
fluxes that are single-peaked.  Colour-colour diagram position is
similar to the 1984 observations: by this measure too there
are single-peaked bursts at higher accretion rates \citep{van90}.

What about the other sources that have exhibited non-PRE double-peaked
bursts? For 4U 1608-52 there is one multi-peak burst in the EXOSAT
sample \citep{pen89} and one in the RXTE catalogue (Burst 24 from this
source, G07 Figure 9).  \citet{kuu02} and \citet{jon04} discovered one
multi-peak burst apiece during RXTE observations of GX 17+2 and 4U
1709-267.    Figure 9 of G07 also reveals previously unreported
multi-peak non PRE bursts from Aql X-1 (Bursts 7 and 32) and EXO
0748-676 (Burst 82). Every single one of these sources has 
single-peaked bursts at higher accretion rates.

\begin{figure}
  \resizebox{\hsize}{!}{\includegraphics{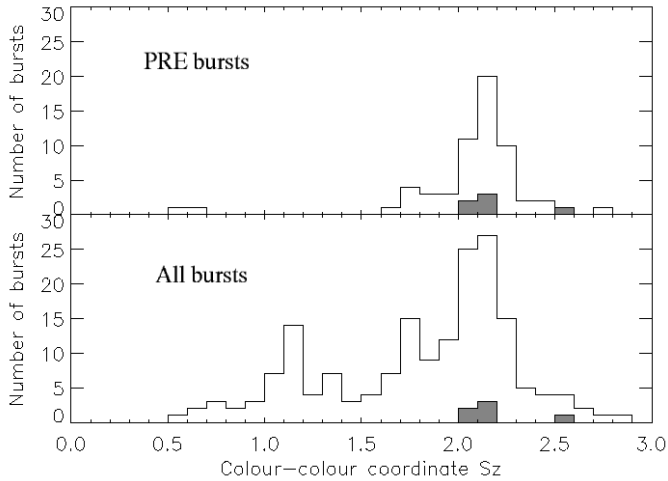}}
  \caption{The distribution of multi-peaked bursts (shaded) with $S_z$
    compared to other bursts for the three sources in the RXTE
    catalogue that have (a) non PRE multipeak bursts and (b) trace out a long
    enough path in the colour-colour diagram to calculate
    $S_z$.  These sources are  
    4U 1636-536, Aql X-1 and 4U 1608-52.  The lower panel shows the
    distribution compared to all other bursts, the upper panel the
    distribution compared to PRE bursts.}
  \label{f3}
\end{figure}

Given this sample, what can we say about the accretion rate at
which multi-peak bursts occur?   Three of the sources in the RXTE
catalogue with multi-peak bursts trace out a full enough path in the
colour-colour diagram to 
allow computation of $S_z$:  4U 1636-536, 4U 1608-52 and Aql X-1
(although for the latter $S_z$ can only be calculated for one of the
two multi-peak bursts).  Figure \ref{f3} compares the distribution
of the multi-peak and single-peaked bursts with
$S_z$ for these three sources. Five of the six bursts cluster around
$S_z = 2.1$, but the 
remaining burst (Burst 4 from Figure \ref{f1}) occurs at $S_z \approx
2.5$.   The multi-peak non PRE bursts do seem to be associated with higher
accretion rates:  a two-sided Kolmogorov-Smirnov test gives only a
4\% probability of them being drawn from the same
underlying distribution as the set of all other bursts.
The distribution of
multi-peak bursts is however completely consistent with the
distribution of PRE bursts, which are also associated with higher
accretion rates. However, the sample of multi-peaked bursts is
small, so we 
should treat these statistics with some caution.

We can also consider the dependence on absolute accretion rate as
inferred from the persistent flux for those sources with a reasonable
estimate of distance.  4U 1636-536 exhibits bursts at accretion rates
 from $2-13$\% $\dot{M}_\mathrm{Edd}$.  Three of the multipeak
bursts (Bursts (1)-(3) in Figure \ref{f1}) occur at accretion rates in
the range 3.2 to 3.7\% $\dot{M}_\mathrm{Edd}$, while the other (Burst
(4) in Figure \ref{f1})  occurs at 8\% $\dot{M}_\mathrm{Edd}$.  4U 1608-52 has
bursts for accretion rates in the range 0.5-10\%
$\dot{M}_\mathrm{Edd}$, with the
multipeak burst occurring at 3\% $\dot{M}_\mathrm{Edd}$.  Aql X-1 shows 
bursts for accretion rates in the range 0.1-11\%
$\dot{M}_\mathrm{Edd}$.  The two 
multipeak bursts are at 2.3\% $\dot{M}_\mathrm{Edd}$ and 3.3\%
$\dot{M}_\mathrm{Edd}$.  EXO
0748-676 has only been seen to burst at much lower accretion rates, in
the range 0.5 - 1.5\%
$\dot{M}_\mathrm{Edd}$.  The multipeak burst occurs at 0.9\% $\dot{M}_\mathrm{Edd}$.  For
4U 1709-267 no calibration of accretion rate is possible as there are
no definite PRE bursts.  The
multipeak burst from GX 17+2 is particularly unusual since this source
is a Z rather than an atoll source and accretes at a rate apparently in excess of 
$\dot{M}_\mathrm{Edd}$.  However, the morphology of this burst does
differ from the other examples. 

\section{Discussion}
\label{models}

The recent modelling by CN07 indicates that polar ignition should only
take place above a certain accretion rate.  If multi-peak bursts are
triggered by polar ignition, as suggested by \citet{bha06a, bha06b},
it follows that they should occur at higher accretion rates than the
single-peaked bursts.  This is clearly not the case.  The bursts of 4U
1636-536 also  
pose one other problem for the polar ignition model.  \citet{bha06a, bha06b} model bursts
(3) and (4) in Figure \ref{f1}.  For burst (3) they infer polar
ignition, while for burst (4) they infer intermediate latitude
ignition.  Within the model of CN07, this would require burst (4) to
appear at lower accretion rates than burst (3).  The data show the
exact opposite:  burst (4) has a higher apparent accretion rate than 
bursts (1)-(3).   So where does this leave the polar ignition model?  

\subsection{Are multiple peaks a unique signature of polar ignition?}

The polar ignition model could still be viable if the majority of
polar ignition bursts are single-peaked.  In this scenario all of the
bursts of 4U 1636-536 for 
$S_z \gtrsim 2$ would be polar ignition bursts.   If this were the
case, there are several issues to be resolved.  Firstly, this would
set the accretion rate for the transition from equatorial to polar
ignition at a few percent of $\dot{M}_\mathrm{Edd}$ (G07, and Figure \ref{f2}).
The modelling of CN07 suggests however that this limit should be
higher:  in an example with a neutron star spinning at 650 Hz, CN07
find a transition accretion rate of $\approx 16$\%
$\dot{M}_\mathrm{Edd}$.  The spin rate of 4U 1636-536 is lower, at 580
Hz \citep{str98}, which would push the expected transition rate up.
This disparity would need to be resolved. However, as we have
discussed 
above, precise calibration of $S_z$ and accretion rate is difficult,
and there are other areas of burst physics where similar 
discrepancies arise.  

\begin{figure}
  \resizebox{\hsize}{!}{\includegraphics{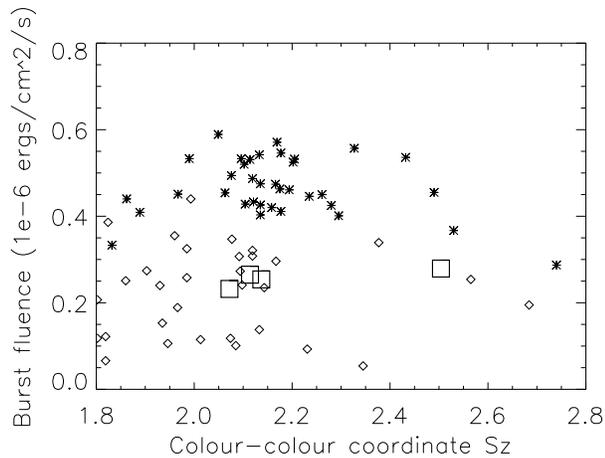}}
  \caption{Burst fluence against accretion rate as measured by $S_z$
    (data taken from G07). Diamonds: single peaked non-PRE bursts;
    Asterisks: PRE bursts; Squares:  multi-peak non PRE bursts.
    Error bars are similar in size to the diamonds.}
  \label{f4}
\end{figure}

We must then consider the reason why a few,
but not the vast majority, of the bursts are multi-peaked:  the stalling
mechanism.  \citet{bha06a, bha06b}, revisiting an idea originally
suggested by \citet{szt85}, proposed that stalling is caused by the
interaction of 
the burning front with a spreading layer of accreted
material \citep{ino99}.  They suggest that stronger bursts may be able
to disrupt this spreading layer and overcome the stalling mechanism,
giving rise to single-peaked bursts.  If this is the case, we would expect the fluence of the
multi-peak bursts to be lower than that of other bursts in the same
region of the colour-colour diagram.  Figure \ref{f4} shows that
although the multi-peak bursts do have low fluences, there are
single-peaked bursts at similar accretion rates with fluences that are
the same or even lower. For the polar ignition model to survive it would be
necessary to explain these low fluence single-peaked bursts.  One
possibility might be that these lower fluence bursts are 
so weak that the burning front stalls completely at the equator and
never reaches the second hemisphere at all.

\subsection{Do other factors affect the latitude dependence
  of ignition?}

The other possibility is that the models of CN07 do not fully
characterize the problem.   As the authors themselves note,
whilst their models do reproduce, for example, average burst
recurrence times, they do not necessarily reproduce the scatter around
this average.  There are clearly subtleties waiting to be resolved
(see also the discussion in \citet{coo06}).  CN07 assume, for
example, that variations in the conditions on the 
stellar surface are set only by the effective gravitational
potential.   We can break this assumption down into several elements.
Firstly, that accreted fuel spreads rapidly and is
not impeded in any way.  Secondly that a similar
re-adjustment occurs after a burst if any unburnt fuel remains.  
Thirdly, that there is no additional asymmetry that would affect burst
triggering.  If any of these conditions were violated, they might
permit polar and equatorial ignition to co-exist at the same accretion
rate.  

The assumption of fuel
spread, based on estimates of its viscosity, is hard to dispute
\citep{bil98}.  Magnetic confinement could lead to asymmetries
in fuel distribution if the field were strong enough.  However, there
is no evidence for persistent pulsations in 4U 1636-536, suggesting
that the magnetic field is in this case too weak to channel or confine
fuel.  What about additional sources of latitude-dependent asymmetry?
Two effects that 
should perhaps be considered are (a) the effect on ignition conditions
(not just stalling conditions)
of a spreading layer at the equator
\citep{ino99}, 
and (b) latitude-dependent temperature asymmetries.  The latter could
arise if, for example, burning fronts 
launched from the equator repeatedly stalled when approaching the pole (due
to reduced flame speed, \citet{spi02}).  Reduced heat injection into
the crust at high latitudes could lead to a temperature gradient that
might affect subsequent ignition conditions.  Latitude dependent
heating due to either core fluid oscillations or slightly asymmetric
accretion could also play a role. We note that both of these
mechanisms have been discussed as a means of generating a quadrupole
moment that 
might limit the spin of rapidly rotating neutron stars via
gravitational wave emission \citep{ush00}.  

\subsection{Alternatives to the polar ignition model}

If the polar ignition model cannot be reconciled with the
observations, an alternative is required.  The nuclear waiting point
model of \citet{fis04} 
predicts multi-peak  
bursts at accretion rates of a few percent $\dot{M}_\mathrm{Edd}$, in
line with the observations. The existence of multiple waiting points in the rp
process also offers a natural explanation for the triple peaked burst,
something that is hard to explain in the polar ignition model.
The alternative is two stage burning
triggered by hydrodynamical instabilities 
\citep{fuj88}.  \citet{bha06a} criticised this model on the grounds
that it was hard to see how an unburnt layer of fuel could be
maintained for long enough for the mechanism to work.  However, a
recent study by \citet{gal06} of the PRE bursts 
of 4U 1636-536 suggested the presence of an additional source of
stratification in this star:  this may promote the survival
of unburnt fuel layers. Theoretical effort is required, however, to generate models of spectral evolution and timing behaviour
for these alternative mechanisms in order to subject them to rigorous
test.   

\section{Conclusions}

We have presented an analysis of the accretion rate dependence of
multi-peaked non PRE bursts, focusing in particular on whether they
occur at higher accretion rates than single-peaked bursts.  The fact
that they do not contradicts one of the
predictions of the polar ignition model, given the recent study by CN07
that predicts a strong dependence of ignition latitude on
accretion rate.  While there are ways in which the polar
ignition/stalling model could be reconciled with the data, alternative
models should also be pursued.  Ultimately a combination of models may prove to
be the answer, particularly given that the multi-peak bursts are not a
homogeneous set.  To see this one only has to consider
the differences in morphology and apparent accretion rate for the four
bursts in Figure \ref{f1}.

\acknowledgements

We thank Lars Bildsten, Tod Strohmayer, Randall Cooper, Henk Spruit,
Rashid Sunyaev, 
Duncan Galloway and Stuart Sim for comments.

\end{document}